# Ubiquity of Fourier Transformation in Optical Sciences (Part II)

Masud Mansuripur
James C. Wyant College of Optical Sciences, The University of Arizona, Tucson

**Abstract**. This paper contains a transcript of my presentation at the *Wyant Tribute Symposium* on August 2, 2021 at SPIE's Optics & Photonics conference in San Diego, California. The technical part of the paper has no overlap with a previous article of mine that was published in *Applied Optics* last year, bearing the same title as this one.[1] The applications of Fourier transformation described in the present paper include the central limit theorem of probability and statistics, the Shannon-Nyquist sampling theorem, and computing the electromagnetic field radiated by an oscillating magnetic dipole.

**1. Introduction**. Let me begin by saying that I have known Jim for 33 years. This is what he looked like when I joined the Optical Sciences Center (now College of Optical Sciences) in the Fall of 1988. You can tell that he hasn't changed a bit.

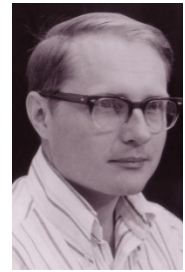

In contrast, see how the rest of us have aged; this is what *I* looked like in those days!

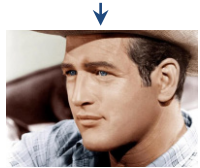

For reasons that will become clear in a few minutes, whenever I think of Jim Wyant, the following story pops into my head.

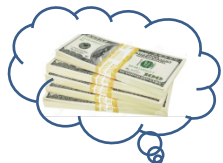
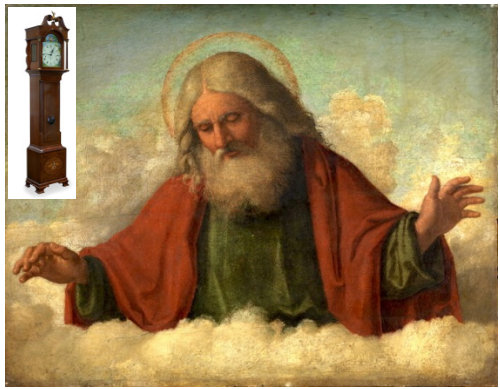
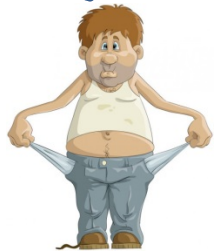

A destitute man happens to meet the Lord Almighty.
He asks: "God, what's a million years to you?"
The Lord responds: "Oh, nothing; it's just like a second to me."
The poor man asks: "God, what's a million dollars to you?"
The Lord says: "Nothing; it's like a penny to me."
The man finally says: "Dear Lord, may I have a penny?"
To which the Lord responds: "Absolutely; please wait a second!"

Now, here's the background story. There was a time that I thought I had many rich and generous friends. So, I had this idea of asking each of them to donate half a million dollars to our College, so that we could create endowed scholarships in their own names. I asked Jim to accompany me to the office of the University's vice-president to see if she would agree to match such donations by covering the students' tuition and other expenses. To our delight, vice-president Leslie Tolbert loved the idea and consented to our request. Moreover, she allowed us to have as many as 30 such endowed scholarships with matching provided from her office's budget.

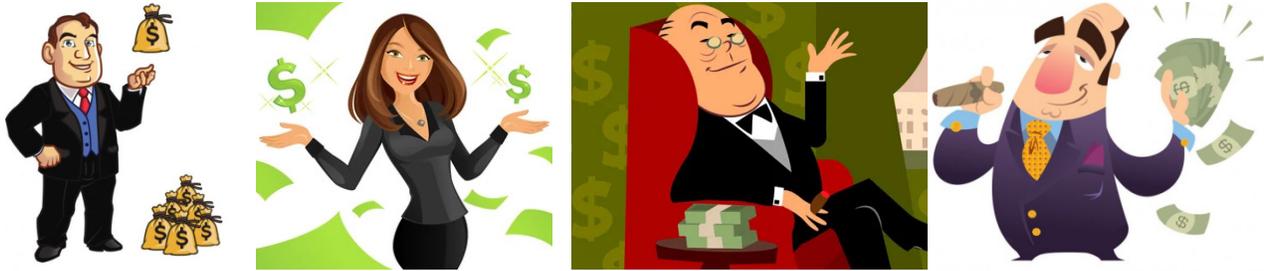

I have many rich and generous friends!

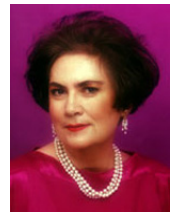

On our way back from the VP's office, I turned to Jim and said: "Jim, why don't *you* become a donor and give us the first half-million dollars to start this *Friends of Tucson Optics* (FoTO) scholarship program. To which Jim responded: "You'll never get a penny from me!" Fortunately for our College and for our students (then and forever after), we didn't have to wait a million years for Jim to change his mind! In less than two weeks, Jim had decided to endow the first FoTO scholarship — to be named after his beloved deceased wife, Louise.

**Louise Wyant**

We went on to receive a few more gifts from generous individuals and established a couple more endowed FoTO scholarships. A few years later, Jim announced that he would donate another 10 million dollars to the College, provided that the College would use the money for a four-to-one matching of $100,000 gifts. That is when the money started pouring in and, in a matter of two to three years, we had all thirty FoTO scholarships fully endowed, thanks to Jim's generosity.

As for my rich and generous friends that I had initially intended to rely upon, it turned out that my rich friends weren't generous, and my generous friends weren't rich!

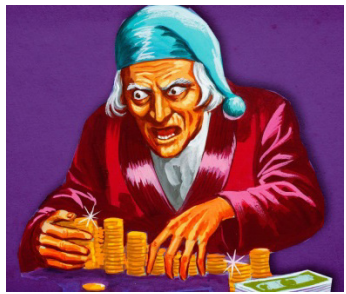

My rich friends aren't generous,
and my generous friends aren't rich!

With regard to my presentation today, I wanted to do something that touched upon the educational aspects of our College. As you all know, Jim is a gifted teacher and, to my eternal consternation (!), our students have always had a much higher opinion of him as an instructor than they've had of the rest of us on the faculty. So, I thought I should contribute a paper to this symposium that might have some educational value. That is when I hit upon the unifying theme of Fourier transformation.

Our colleague, Prof. Sasian, edited a special edition of *Applied Optics* last year. This was a commemorative issue in honor of the renaming of our College to *James C. Wyant College of*

*Optical Sciences*. My plan was to submit a paper entitled "Ubiquity of Fourier Transformation in Optical Sciences" to that special issue of *Applied Optics*, then present some of its results at this Tribute Symposium to Dean Wyant, which was initially scheduled for last August. Unfortunately, the pandemic intervened and the Tribute was postponed until this week.

That *Applied Optics* paper[1] has now been in the public domain for more than a year and, in any event, there isn't much time available today for me to discuss the technical aspects of the paper in any detail. So, I'll just mention the kind of topics that are covered in the paper, hoping that you will find them sufficiently interesting to look up the details on your own. The published version of this presentation, however, will contain a few more applications of Fourier transformation that were not covered in the *Applied Optics* paper.

The organization of this paper is as follows. In Sec.2, I describe the central limit theorem of probability and statistics, which is the fundamental mathematical argument as to why many natural phenomena appear to have a Gaussian probability distribution. The comb function, comprising an infinite array of identical, equi-spaced delta-functions, is the subject of Sec.3, where I use two different methods to show that $\text{comb}(s)$ is the Fourier transform of $\text{comb}(x)$. Section 4 describes how one may use the comb function to derive the Fourier series coefficients of periodic functions. Another application of the comb function in the context of the Shannon-Nyquist sampling theorem is the subject of Sec.5. Finally, in Sec.6, I use a Fourier-transform-based method to solve Maxwell's equations of classical electrodynamics and derive the radiation field of a magnetic dipole oscillator.

**2. The central limit theorem**. It is a well-known fact that the Fourier transform of the unit rectangular function, $\text{rect}(x)$, is $\text{sinc}(s) = \sin(\pi s)/(\pi s)$.[2] Successive convolutions of $\text{rect}(x)$ with itself yield $\text{tri}(x) = \text{rect}(x) * \text{rect}(x)$, whose Fourier transform is $\text{sinc}^2(s)$, and $\text{hump}(x) = \text{rect}(x) * \text{rect}(x) * \text{rect}(x)$, whose Fourier transform is $\text{sinc}^3(s)$.[†] Clearly, the $N$-fold convolution of $\text{rect}(x)$ with itself yields a function of total width $N$, unit area, and Fourier transform $\text{sinc}^N(s)$, whose profile along the $x$-axis approaches that of a Gaussian function as $N$ continues to rise.

Considering that $\sin(\pi s) = \pi s - (\pi s)^3/3! + \cdots$, it is seen that $\text{sinc}(s) \cong 1 - \tfrac{1}{6}(\pi s)^2$ in the immediate vicinity of $s = 0$ and that, further away from the origin, it has positive as well as negative values whose magnitudes are significantly smaller than 1. Thus, $\text{sinc}^N(s)$ for large $N$ is a function of $s$ that, to a good approximation, equals $[1 - \tfrac{1}{6}(\pi s)^2]^N$ in the immediate neighborhood of the origin, and remains close to zero elsewhere. Given that $e^{-\alpha s^2} \cong 1 - \alpha s^2$ around $s = 0$, we conclude that $\text{sinc}^N(s)$ approaches the Gaussian function $e^{-(N\pi^2/6)s^2}$ for sufficiently large $N$. The inverse Fourier transform of this function, namely, $f(x) = \sqrt{6/(N\pi)} \exp(-6x^2/N)$, should thus provide a good approximation to the $N$-fold convolution of $\text{rect}(x)$ with itself.

The method described in the preceding paragraph is quite general and can be applied to the convolution of a large number of functions that may or may not be identical, say, $f_1(x) * f_2(x) * \cdots * f_N(x)$, provided that the individual functions satisfy certain constraints. As an example, let us consider a large number of independent random variables, say, $x_1, x_2, \cdots, x_N$, having respective probability distributions $p_1(x), p_2(x), \cdots, p_N(x)$. In general, $p_n(x) \geq 0$ and $\int_{-\infty}^{\infty} p_n(x) \mathrm{d}x = 1$. The sum $x_{\text{sum}} = x_1 + x_2 + \cdots + x_N$ of these random variables is known to have the probability distribution $p_{\text{sum}}(x) = p_1(x) * p_2(x) * \cdots * p_N(x)$.[3]

---

[†]Here, we are using the standard definition of $\text{rect}(x) = 1$ when $|x| \leq \tfrac{1}{2}$, and 0 otherwise. Similarly, $\text{tri}(x) = 1 - |x|$ when $|x| \leq 1$, and 0 otherwise.[2] The name used for the function $\text{hump}(x) = \tfrac{3}{4} - x^2$ when $|x| \leq \tfrac{1}{2}$, and $\tfrac{1}{2}(|x| - \tfrac{3}{2})^2$ when $\tfrac{1}{2} \leq |x| \leq \tfrac{3}{2}$, and 0 when $|x| \geq \tfrac{3}{2}$, is *not* standard. The function $\text{sinc}(x)$ is often defined as $\sin(\pi x)/(\pi x)$.

Next, we define an average value $\bar{x}_n = \int_{-\infty}^{\infty} x p_n(x) dx$ and variance $\sigma_n^2 = \int_{-\infty}^{\infty}(x - \bar{x}_n)^2 p_n(x) dx$ for each random variable, then shift $p_n(x)$ by $\bar{x}_n$ along the $x$-axis to arrive at the shifted (or centered) probability-density function $p_{nc}(x) = p_n(x + \bar{x}_n)$, which satisfies the following identities:

$$\int_{-\infty}^{\infty} x p_{nc}(x) dx = \int_{-\infty}^{\infty} x p_n(x + \bar{x}_n) dx = \int_{-\infty}^{\infty} (x - \bar{x}_n) p_n(x) dx = \bar{x}_n - \bar{x}_n \int_{-\infty}^{\infty} p_n(x) dx = 0. \quad (1)$$

$$\int_{-\infty}^{\infty} x^2 p_{nc}(x) dx = \int_{-\infty}^{\infty} x^2 p_n(x + \bar{x}_n) dx = \int_{-\infty}^{\infty} (x - \bar{x}_n)^2 p_n(x) dx = \sigma_n^2. \quad (2)$$

Denoting the so-called characteristic function[2-4] of the probability density $p(x)$ by $\tilde{p}(s) = \int_{-\infty}^{\infty} p(x) e^{i2\pi sx} dx$, we find the following identities among $\bar{x}_n$, $\sigma_n$, $p_n(x)$, $\tilde{p}_n(s)$, $\tilde{p}_{nc}(s)$ and $\tilde{p}_{sum}(s)$:

$$\tilde{p}_n(s) = \int_{-\infty}^{\infty} p_n(x) e^{i2\pi sx} dx = \int_{-\infty}^{\infty} p_n(x + \bar{x}_n) e^{i2\pi s(x + \bar{x}_n)} dx = \exp(i2\pi \bar{x}_n s) \tilde{p}_{nc}(s). \quad (3)$$

$$d\tilde{p}_n(s)/ds|_{s=0} = i2\pi \bar{x}_n. \quad (4)$$

$$d\tilde{p}_{nc}(s)/ds|_{s=0} = 0. \quad (5)$$

$$d^2 \tilde{p}_{nc}(s)/ds^2|_{s=0} = -4\pi^2 \sigma_n^2. \quad (6)$$

$$\tilde{p}_{sum}(s) = \prod_{n=1}^{N} \tilde{p}_n(s) = \exp(i2\pi \sum_{n=1}^{N} \bar{x}_n s) \prod_{n=1}^{N} \tilde{p}_{nc}(s). \quad (7)$$

It is now easy to verify that, for the $N$ independent random variables under consideration, $\bar{x}_{sum} = \sum_{n=1}^{N} \bar{x}_n$ and $\sigma_{sum}^2 = \sum_{n=1}^{N} \sigma_n^2$. Moreover, given that $\tilde{p}_{nc}(s) \cong 1 - 2\pi^2 \sigma_n^2 s^2 \cong e^{-2\pi^2 \sigma_n^2 s^2}$ in the immediate vicinity of $s = 0$, if all characteristic functions $\tilde{p}_{nc}(s)$ happen to have a magnitude strictly less than 1.0 at $s \neq 0$, their product $\prod_{n=1}^{N} \tilde{p}_{nc}(s)$ will rapidly decline toward zero away from the origin ($s = 0$). One can then approximate Eq.(7) as follows:

$$\tilde{p}_{sum}(s) \cong \exp(i2\pi \sum_{n=1}^{N} \bar{x}_n s) \prod_{n=1}^{N} [1 - 2\pi^2 \sigma_n^2 s^2] \cong \exp(i2\pi \bar{x}_{sum} s) \exp(-2\pi^2 \sigma_{sum}^2 s^2). \quad (8)$$

The inverse Fourier transform of Eq.(8) now yields the probability density function of $x_{sum}$ as

$$p_{sum}(x) = (\sqrt{2\pi} \sigma_{sum})^{-1} \exp[-(x - \bar{x}_{sum})^2/(2\sigma_{sum}^2)]. \quad (9)$$

This is the main conclusion of the central limit theorem of probability and statistics, explaining why so many natural phenomena, each arising as the superposition of a large number of more or less independent random variables, tend to have a Gaussian probability distribution.

**3. The comb function**. The function $comb(x)$ is a periodic function of the real variable $x$, where each period consists of a single delta-function; that is, $comb(x) = \sum_{n=-\infty}^{\infty} \delta(x - n)$. The standard way of showing that the Fourier transform of $comb(x)$ is $comb(s)$ starts by truncating the tails of the function where $|n| > N$, evaluating the Fourier transform of the truncated function, then allowing $N$ to rise to infinity.[2,4] We thus find

$$\mathcal{F}\{\sum_{n=-N}^{N} \delta(x - n)\} = \sum_{n=-N}^{N} \int_{-\infty}^{\infty} \delta(x - n) e^{-i2\pi sx} dx = \sum_{n=-N}^{N} e^{-i2\pi ns}$$

$$= e^{i2\pi Ns} \sum_{n=0}^{2N} e^{-i2\pi ns} = e^{i2\pi Ns} [1 - e^{-i2\pi(2N+1)s}]/(1 - e^{-i2\pi s})$$

$$= \sin[\pi(2N+1)s]/\sin(\pi s). \quad (10)$$

Now, in the limit when $N \to \infty$, the function $\sin[\pi(2N+1)s]$ oscillates rapidly with a changing $s$, and one must carefully examine its behavior in the vicinity of $s_n = n\pi$ (i.e., around the points where $s$ is an integer multiple of $\pi$), because the denominator $\sin(\pi s)$ as well as the numerator are very close to zero around such points. It is not difficult to see that, in the immediate vicinity of each $s_n$, the function approaches $(2N+1)\text{sinc}[(2N+1)(s-s_n)]$, where, as before, the sinc function is defined as $\text{sinc}(s) = \sin(\pi s)/(\pi s)$. Thus, in the limit when $N \to \infty$, the function appearing on the right-hand side of Eq.(1) approaches $\text{comb}(s)$.

An alternative way of demonstrating the same thing involves raising the function $\cos(\pi x)$ to the power $2N$, where $N$ is a positive integer, then allowing $N$ to go to infinity. In the immediate vicinity of $x_n = n$, where $n$ is a positive, zero, or negative integer, the function $\cos^2(\pi x)$ is very close to 1, which makes its $N^{\text{th}}$ power also close to 1. However, as soon as $x$ moves away from $x_n$, $\cos^2(\pi x)$ drops below zero and its $N^{\text{th}}$ power rapidly approaches zero as $N \to \infty$. Thus, in $\cos^{2N}(\pi x)$, we have a periodic function consisting if narrow bumps of height 1, centered at $x_n = n$, which approaches zero everywhere else when $N \to \infty$. The area under each bump is readily found using the method of integration by parts, as follows:

$$\int_{-\frac{1}{2}}^{\frac{1}{2}} \cos^{2N}(\pi x)\,dx = \int_{-\frac{1}{2}}^{\frac{1}{2}} [1 - \sin^2(\pi x)]\cos^{2(N-1)}(\pi x)\,dx$$

$$= \int_{-\frac{1}{2}}^{\frac{1}{2}} \cos^{2(N-1)}(\pi x)\,dx + \left.\frac{\sin(\pi x)\cos^{2N-1}(\pi x)}{(2N-1)\pi}\right|_{x=-\frac{1}{2}}^{\frac{1}{2}} - \frac{1}{2N-1}\int_{-\frac{1}{2}}^{\frac{1}{2}} \cos^{2N}(\pi x)\,dx$$

$$\to \int_{-\frac{1}{2}}^{\frac{1}{2}} \cos^{2N}(\pi x)\,dx = \left(\frac{2N-1}{2N}\right)\int_{-\frac{1}{2}}^{\frac{1}{2}} \cos^{2(N-1)}(\pi x)\,dx. \tag{11}$$

Considering that $\int_{-\frac{1}{2}}^{\frac{1}{2}} dx = 1$, Eq.(11) yields $\int_{-\frac{1}{2}}^{\frac{1}{2}} \cos^{2N}(\pi x)\,dx = (2N-1)!!/(2N)!!$. This area, of course, approaches zero as $N \to \infty$, because the bumps become narrower and narrower while their height remains fixed at 1. However, the function $(2N)!!\cos^{2N}(\pi x)/(2N-1)!!$, whose bumps continually increase in height as their width shrinks with an increasing $N$, properly represent the function $\text{comb}(x)$ in the limit when $N \to \infty$.

Next, we consider the Fourier transform of $f(x) = \cos^2(\pi x) = \frac{1}{2} + \frac{1}{4}e^{i2\pi x} + \frac{1}{4}e^{-i2\pi x}$, namely, $F(s) = \frac{1}{2}\delta(s) + \frac{1}{4}\delta(s-1) + \frac{1}{4}\delta(s+1)$. Using the convolution theorem of Fourier transformation,[2,4] the following pattern emerges:

$$\mathcal{F}\{\cos^2(\pi x)\} = F(s) = \frac{1}{2^2}[\delta(s+1) + 2\delta(s) + \delta(s-1)]$$

$$\mathcal{F}\{\cos^4(\pi x)\} = F(s) * F(s) = \frac{1}{2^4}[\delta(s+2) + 4\delta(s+1) + 6\delta(s) + 4\delta(s-1) + \delta(s-2)]$$

$$\vdots$$

$$\mathcal{F}\{\cos^{2N}(\pi x)\} = \overbrace{F(s) * F(s) * \cdots * F(s)}^{N \text{ times}} = \frac{1}{2^{2N}}\sum_{n=0}^{2N}\binom{2N}{n}\delta(s+N-n). \tag{12}$$

This is an array of $\delta$-functions centered at $s = 0$ and extending (symmetrically) over integer values of $s$ from $-N$ to $N$. The magnitude of the largest $\delta$-function, located at $s = 0$, is $\binom{2N}{N}/2^{2N}$. Scaling of $\cos^{2N}(\pi x)$ by the factor $(2N)!!/(2N-1)!!$ causes the magnitude of this central $\delta$-function to become

$$\frac{(2N)!!}{(2N-1)!!} \times \frac{(2N)!}{2^{2N}(N!)(N!)} = \frac{2^N(N!)}{(2N-1)!!} \times \frac{(2N-1)!! \times 2^N(N!)}{2^{2N}(N!)(N!)} = 1. \tag{13}$$

The adjacent δ-functions located at $s = \pm 1$ are slightly smaller, having magnitude $N/(N+1)$. The next adjacent δ-functions located at $s = \pm 2$ are a bit smaller, at $N(N-1)/(N+1)(N+2)$, and so on. It should thus be clear that, as $N \to \infty$ and the function $(2N)!! \cos^{2N}(\pi x)/(2N-1)!!$ approaches $\text{comb}(x)$, its Fourier transform, a tapered array of $2N+1$ delta-functions centered at $s = 0$, approaches $\text{comb}(s)$.

**4. The Fourier series**. Many textbooks on Fourier transform theory begin by introducing the Fourier series, associated with periodic functions, as a first step toward discussing the Fourier integral for an arbitrary function $f(x)$. However, the reverse approach of starting with the Fourier integral $F(s) = \int_{-\infty}^{\infty} f(x) e^{-i2\pi sx} dx$, then specializing to the case of periodic functions, has certain merits. Let $f_p(x)$ represent a single period, confined to the interval $(x_0, x_0 + p]$, of the periodic function $f(x)$. Recalling that $p^{-1}\text{comb}(x/p)$ is a periodic array of unit δ-functions located at $x_n = np$, where the integer $n$ could be positive, zero, or negative, one may write $f(x) = p^{-1}\text{comb}(x/p) * f_p(x)$. The convolution theorem[2,4] now yields

$$F(s) = \text{comb}(ps)F_p(s) = p^{-1} \sum_{n=-\infty}^{\infty} F_p(n/p)\delta[s - (n/p)]. \tag{14}$$

Thus, the Fourier transform of a periodic function consists of an array of equi-spaced δ-functions (located at $s_n = n/p$), where the amplitude of the $n^{\text{th}}$ delta-function is given by

$$p^{-1}F_p(n/p) = p^{-1} \int_{x_0}^{x_0+p} f_p(x) \exp(-i2\pi nx/p) \, dx. \tag{15}$$

The inverse Fourier transform now yields the Fourier series of the periodic function, as follows:

$$f(x) = \mathcal{F}^{-1}\{F(s)\} = \int_{-\infty}^{\infty} p^{-1} \sum_{n=-\infty}^{\infty} F_p(n/p)\delta[s-(n/p)] e^{i2\pi sx} ds$$

$$= p^{-1} \sum_{n=-\infty}^{\infty} F_p(n/p) \exp(i2\pi nx/p). \tag{16}$$

In the special case when $f_p(x)$ is real-valued, Eq.(15) yields $F_p(-n/p) = F_p^*(n/p)$. Consequently,

$$f(x) = p^{-1}F_p(0) + 2p^{-1} \sum_{n=1}^{\infty} \{\text{Re}[F_p(n/p)] \cos(2\pi nx/p) - \text{Im}[F_p(n/p)] \sin(2\pi nx/p)\}$$

$$= a_0 + \sum_{n=1}^{\infty} a_n \cos(2\pi nx/p) + \sum_{n=1}^{\infty} b_n \sin(2\pi nx/p). \tag{17}$$

In the above equation,

$$a_0 = p^{-1}F_p(0) = p^{-1} \int_{x_0}^{x_0+p} f_p(x) dx, \tag{17a}$$

$$a_n = 2p^{-1}\text{Re}[F_p(n/p)] = 2p^{-1} \int_{x_0}^{x_0+p} f_p(x) \cos(2\pi nx/p) \, dx, \tag{17b}$$

$$b_n = -2p^{-1}\text{Im}[F_p(n/p)] = 2p^{-1} \int_{x_0}^{x_0+p} f_p(x) \sin(2\pi nx/p) \, dx. \tag{17c}$$

In this way, the Fourier series coefficients of a periodic function $f(x)$ of period $p$, are obtained as the normalized sampled values $p^{-1}F_p(n/p)$ of the Fourier transform $F_p(s)$ of a single period $f_p(x)$, in accordance with Eq.(16). Alternatively, if $f(x)$ happens to be real-valued, the Fourier coefficients are derived from the sine and cosine transforms of $f_p(x)$, in accordance with Eqs.(17).

**5. The Shannon-Nyquist Sampling theorem.** In modern digital signal processing, electronic communication, and information storage, analog waveforms are sampled at regular intervals, then converted to a digital format.[2,4] The Fourier transform $F(s)$ of the waveform (or signal) under consideration, namely, $f(t)$, is said to be bandlimited if $F(s)$ turns out to be negligible outside the frequency interval $(-w, w)$. Under such circumstances, multiplying $f(t)$ with the sampling function $p^{-1}\text{comb}(t/p)$, namely, an array of unit $\delta$-functions at regular intervals $\Delta t = p$, creates a sampled function $p^{-1}\text{comb}(t/p)f(t)$, whose Fourier transform, $\text{comb}(ps) * F(s)$, may be written as $p^{-1}\sum_{n=-\infty}^{\infty} F[s - (n/p)]$. This sum over the shifted copies of $F(s)$ by integer-multiples of $1/p$ would preserve the information content of the original signal $f(t)$ if the shifted copies of $F(s)$ do not overlap each other; that is, if $1/p > 2w$. In other words, the rate $1/p$ of sampling $f(t)$ at regular intervals $p$ must exceed the (two-sided) width $2w$ of the frequency spectrum $F(s)$ of $f(t)$.

During the reconstruction of the original signal $f(t)$ from its uniformly-spaced samples $f(np)$, the discretized function $p^{-1}\text{comb}(t/p)f(t)$ is sent through a linear, shift-invariant, low-pass filter, whose transfer function $H(s) = p\,\text{rect}(ps)$ multiplies the Fourier transform of the input waveform. Assuming the sampling rate satisfies the aforesaid Nyquist criterion, the individual copies of $F(s)$ that are present within the spectrum of the input waveform will be well-separated from each other, thus allowing the filter to extract the original (unshifted) copy of $F(s)$ and perfectly reproduce the signal $f(t)$ at its output. The inverse Fourier transform of $H(s)$ is the so-called impulse-response of the filter, which is presently given by $h(t) = \text{sinc}(t/p)$. The output $f(t)$ of the filter, being the convolution between the input waveform and the impulse-response, is thus seen to be

$$f(t) = \sum_{n=-\infty}^{\infty} f(np)\text{sinc}[(t - np)/p]. \tag{18}$$

The function $\text{sinc}(t/p)$, whose properly shifted and scaled copies reconstruct the original signal $f(t)$ in accordance with Eq.(18), is known as the interpolation function.[2]

**6. Radiation by an oscillating magnetic point-dipole.** Consider a magnetic point-dipole $m_0\hat{z}$, sitting at the origin of the $xyz$ coordinates in free space and oscillating at the fixed frequency $\omega_0$. The spatio-temporal magnetization distribution of this dipole is written as

$$\boldsymbol{M}(\boldsymbol{r}, t) = m_0\delta(x)\delta(y)\delta(z)\cos(\omega_0 t)\,\hat{z}. \tag{19}$$

Fourier transforming the above magnetization profile in four-dimensional spacetime $(\boldsymbol{r}, t)$, we find

$$\widetilde{\boldsymbol{M}}(\boldsymbol{k}, \omega) = \int_{-\infty}^{\infty} \boldsymbol{M}(\boldsymbol{r}, t)e^{-\mathrm{i}(\boldsymbol{k}\cdot\boldsymbol{r} - \omega t)}\mathrm{d}\boldsymbol{r}\mathrm{d}t = \pi m_0[\delta(\omega + \omega_0) + \delta(\omega - \omega_0)]\hat{z}. \tag{20}$$

(with $\mathrm{d}\boldsymbol{r} = \mathrm{d}x\mathrm{d}y\mathrm{d}z$)

The electric current-density associated with our point-dipole is $\boldsymbol{J}(\boldsymbol{r}, t) = \mu_0^{-1}\boldsymbol{\nabla}\times\boldsymbol{M}(\boldsymbol{r}, t)$; here, $\mu_0$ is the permeability of free space.[5] The Fourier transform $\widetilde{\boldsymbol{J}}(\boldsymbol{k}, \omega) = \mathrm{i}\mu_0^{-1}\boldsymbol{k}\times\widetilde{\boldsymbol{M}}(\boldsymbol{k}, \omega)$ of this current-density now yields the Fourier transform of the vector potential $\boldsymbol{A}(\boldsymbol{r}, t)$ produced (in the Lorenz gauge) by the oscillating magnetic dipole in its surrounding space, as follows:[1,5]

$$\widetilde{\boldsymbol{A}}(\boldsymbol{k}, \omega) = \mu_0\widetilde{\boldsymbol{J}}(\boldsymbol{k}, \omega)/[k^2 - (\omega/c)^2] = \mathrm{i}\boldsymbol{k}\times\widetilde{\boldsymbol{M}}(\boldsymbol{k}, \omega)/[k^2 - (\omega/c)^2]. \tag{21}$$

As for the scalar potential $\psi(\boldsymbol{r}, t)$ produced by the dipole (again, in the Lorenz gauge), we note that, since magnetic dipoles do not possess an electric charge, that is, $\rho(\boldsymbol{r}, t) = 0$, their scalar potential $\widetilde{\psi}(\boldsymbol{k}, \omega) = \varepsilon_0^{-1}\widetilde{\rho}(\boldsymbol{k}, \omega)/[k^2 - (\omega/c)^2]$ is zero; here, $\varepsilon_0$ is the permittivity of free space.[5]

Our next task is to determine the spatio-temporal distribution $\boldsymbol{A}(\boldsymbol{r}, t)$ of the vector potential by inverse Fourier transforming $\widetilde{\boldsymbol{A}}(\boldsymbol{k}, \omega)$ as given by Eq.(21). To this end, we write

$$\begin{aligned}
A(r,t) &= (2\pi)^{-4} \int_{-\infty}^{\infty} \widetilde{A}(k,\omega) e^{i(k\cdot r - \omega t)} dk d\omega \quad \leftarrow \boxed{dk = dk_x dk_y dk_z} \\
&= (2\pi)^{-4} \int_{-\infty}^{\infty} \frac{ik \times \pi m_0 [\delta(\omega+\omega_0)+\delta(\omega-\omega_0)]\hat{z}}{k^2 - (\omega/c)^2} e^{i(k\cdot r - \omega t)} dk d\omega \\
&= -\frac{im_0 \cos(\omega_0 t)\hat{z}}{8\pi^3} \times \int_{-\infty}^{\infty} \frac{k}{k^2-(\omega_0/c)^2} e^{ik\cdot r} dk \\
&\qquad\qquad\qquad\qquad\qquad\qquad \boxed{2i[\sin(kr) - kr\cos(kr)]/(kr)^2} \\
&= -\frac{im_0 \cos(\omega_0 t)\hat{z}}{8\pi^3} \times \int_{k=0}^{\infty} \frac{2\pi k^3 \hat{r}}{k^2-(\omega_0/c)^2} \left(\int_{\vartheta=0}^{\pi} \sin\vartheta \cos\vartheta\, e^{ikr\cos\vartheta} d\vartheta\right) dk \\
&= \frac{m_0 \sin\theta \cos(\omega_0 t)\hat{\varphi}}{2\pi^2 r^2} \int_{k=0}^{\infty} \frac{k[\sin(kr) - kr\cos(kr)]}{k^2-(\omega_0/c)^2} dk \quad \leftarrow \boxed{\hat{z}\times\hat{r} = \sin\theta\,\hat{\varphi}} \\
&= \frac{m_0 \sin\theta \cos(\omega_0 t)\hat{\varphi}}{8\pi^2 r^2} \left\{ \int_{-\infty}^{\infty} \frac{(i-kr)k\exp(-ikr)}{[k-(\omega_0/c)]\times[k+(\omega_0/c)]} dk - \int_{-\infty}^{\infty} \frac{(i+kr)k\exp(ikr)}{[k-(\omega_0/c)]\times[k+(\omega_0/c)]} dk \right\} \\
&\qquad\qquad\qquad \boxed{\text{close the contour in the lower half-plane}} \quad \boxed{\text{close the contour in the upper half-plane}} \\
&= \frac{m_0 \sin\theta \cos(\omega_0 t)\hat{\varphi}}{8\pi r^2} \{[1 + i(\omega_0 r/c)]e^{-i\omega_0 r/c} + [1 - i(\omega_0 r/c)]e^{i\omega_0 r/c}\} \leftarrow \boxed{\text{Cauchy's theorem}} \\
&= \frac{m_0 \sin\theta [\cos(\omega_0 r/c) + (\omega_0 r/c)\sin(\omega_0 r/c)] \cos(\omega_0 t)\hat{\varphi}}{4\pi r^2}. \qquad (22)
\end{aligned}$$

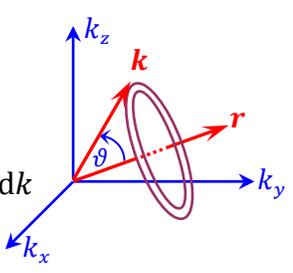

This is neither the retarded nor the advanced vector potential at this point. To arrive at the retarded potential, we need to include an appropriate superposition of plane-waves that exist, independently of our oscillating dipole, in free space. Such plane-waves have a temporal frequency $\omega = \omega_0$ and reside on a spherical surface of radius $k = \omega_0/c$ in the $k$-space.[1] It is not difficult to guess that the desired distribution in the $(k,\omega)$ space of this vacuum vector potential should be

$$\widetilde{A}_{\text{vac}}(k,\omega) = \tfrac{1}{2}\pi^2 m_0 \hat{z} \times \hat{k}\, \delta[k - (\omega_0/c)][\delta(\omega - \omega_0) - \delta(\omega + \omega_0)]. \qquad (23)$$

Here, $\delta(\omega \pm \omega_0)$ ensure that the vacuum potential oscillates at the frequency $\omega_0$; similarly, the confinement of the $k$-vectors to the requisite spherical surface is guaranteed by $\delta[k - (\omega_0/c)]$. The term $\hat{z}\times\hat{k}$ is needed if the plane-waves are to comply with the Lorenz gauge, and also enforces the $\hat{\varphi}$-directed orientation of the vacuum vector potential in the $(r,t)$ space—to be confirmed shortly. Finally, the constant coefficient $\tfrac{1}{2}\pi^2 m_0$ is chosen to make $A_{\text{vac}}(r,t)$ properly combine with our magnetic dipole's vector potential given by Eq.(22). The vacuum vector potential in the $(r,t)$ space is now found straightforwardly by an inverse Fourier transformation of $\widetilde{A}_{\text{vac}}(k,\omega)$ of Eq.(23); that is,

$$\begin{aligned}
A_{\text{vac}}(r,t) &= (2\pi)^{-4} \int_{-\infty}^{\infty} \widetilde{A}_{\text{vac}}(k,\omega) e^{i(k\cdot r - \omega t)} dk d\omega \quad \leftarrow \boxed{dk = dk_x dk_y dk_z} \\
&= (2\pi)^{-4} \int_{-\infty}^{\infty} \tfrac{1}{2}\pi^2 m_0 \hat{z}\times\hat{k}\, \delta[k-(\omega_0/c)][\delta(\omega-\omega_0) - \delta(\omega+\omega_0)] e^{i(k\cdot r - \omega t)} dk d\omega \\
&= -\frac{im_0 \sin(\omega_0 t)\hat{z}}{16\pi^2} \times \int_{k=0}^{\infty} 2\pi k^2 \hat{r}\, \delta[k-(\omega_0/c)]\left(\int_{\vartheta=0}^{\pi} \sin\vartheta \cos\vartheta\, e^{ikr\cos\vartheta} d\vartheta\right) dk \\
&= \frac{m_0 \sin\theta \sin(\omega_0 t)\hat{\varphi}}{4\pi r^2} \int_{k=0}^{\infty} \delta[k-(\omega_0/c)][\sin(kr) - kr\cos(kr)] dk \quad \leftarrow \boxed{\hat{z}\times\hat{r} = \sin\theta\,\hat{\varphi}} \\
&= \frac{m_0 \sin\theta[\sin(\omega_0 r/c) - (\omega_0 r/c)\cos(\omega_0 r/c)]\sin(\omega_0 t)\hat{\varphi}}{4\pi r^2}. \qquad (24)
\end{aligned}$$

Adding the vacuum potential of Eq.(24) to that of the magnetic dipole given by Eq.(22), we finally arrive at the retarded vector potential of the oscillating point-dipole, namely,

$$\boldsymbol{A}_{\text{total}}(\boldsymbol{r},t) = \frac{m_0 \sin\theta}{4\pi}\{r^{-2}\cos[\omega_0(t-r/c)] - (\omega_0/c)r^{-1}\sin[\omega_0(t-r/c)]\}\hat{\boldsymbol{\varphi}}. \qquad (25)$$

This is the exact solution of Maxwell's equations for the vector potential produced by our magnetic point-dipole. Given that the corresponding scalar potential $\psi(\boldsymbol{r},t)$ equals zero, it is easy to verify the satisfaction of the Lorenz gauge condition $\boldsymbol{\nabla}\cdot\boldsymbol{A} + c^{-2}\partial\psi/\partial t = 0$.[5] The electromagnetic fields may now be derived using $\boldsymbol{E}(\boldsymbol{r},t) = -\boldsymbol{\nabla}\psi - \partial\boldsymbol{A}/\partial t$ and $\boldsymbol{B}(\boldsymbol{r},t) = \boldsymbol{\nabla}\times\boldsymbol{A}$, as follows:

$$\boldsymbol{E}(\boldsymbol{r},t) = \frac{m_0 \omega_0 \sin\theta}{4\pi r}\{(\omega_0/c)\cos[\omega_0(t-r/c)] + r^{-1}\sin[\omega_0(t-r/c)]\}\hat{\boldsymbol{\varphi}}. \qquad (26)$$

$$\boldsymbol{B}(\boldsymbol{r},t) = -\frac{m_0 \sin\theta}{4\pi r}(\omega_0/c)^2 \cos[\omega_0(t-r/c)]\,\hat{\boldsymbol{\theta}}$$
$$- \frac{m_0}{4\pi r^2}\{(\omega_0/c)\sin[\omega_0(t-r/c)] - r^{-1}\cos[\omega_0(t-r/c)]\}(2\cos\theta\,\hat{\boldsymbol{r}} + \sin\theta\,\hat{\boldsymbol{\theta}}). \qquad (27)$$

The rate of flow of electromagnetic energy is given by the time-averaged Poynting vector, namely,

$$\langle \boldsymbol{S}(\boldsymbol{r},t)\rangle = \langle \boldsymbol{E}\times\boldsymbol{H}\rangle = \frac{m_0^2 \omega_0^4 \sin^2\theta}{32\pi^2 r^2 Z_0 c^2}\hat{\boldsymbol{r}}. \qquad (28)$$

Here, $Z_0 = (\mu_0/\varepsilon_0)^{\frac{1}{2}} \cong 377\Omega$ is the impedance of free space. Integrating the above rate of flow of energy over a spherical surface of arbitrary radius $r$ yields the total rate of energy radiation by the magnetic dipole, as follows:

$$P = \int_{\theta=0}^{\pi}\langle \boldsymbol{S}(\boldsymbol{r},t)\rangle 2\pi r^2 \sin\theta\,\mathrm{d}\theta = \frac{m_0^2 \omega_0^4}{16\pi Z_0 c^2}\int_{\theta=0}^{\pi}\sin^3\theta\,\mathrm{d}\theta = \frac{m_0^2 \omega_0^4}{12\pi Z_0 c^2}. \qquad (29)^{\ddagger}$$

As expected,[5] the radiated power $P$ is seen to be proportional to the square of the dipole moment $m_0$ and the fourth power of its oscillation frequency $\omega_0$.

**A personal note of gratitude**. I am grateful to Dean James Wyant for, among many other things, providing an environment in which my colleagues and I could thrive, engaging in teaching and research without feeling pressured, and without being burdened by excessive bureaucratic concerns. It has been an honor to be able to call Jim my colleague (and my friend) for the past 33 years.

---

$^{\ddagger}$The convention used in this paper with regard to the magnetic fields is $\boldsymbol{B} = \mu_0\boldsymbol{H} + \boldsymbol{M}$; consequently, the magnetization $\boldsymbol{M}$ has the units of the $B$-field. This means that our magnetic dipole $m_0\hat{\boldsymbol{z}}$, when considered as a small loop of area $\mathcal{A}$ within the $xy$-plane that carries a current $I$ around the $z$-axis, has the dipole moment $m_0 = \mu_0 I\mathcal{A}$. According to Eq.(19), the point-dipole's magnetization $\boldsymbol{M}$ is expressed with the aid of three $\delta$-functions, each having the dimensions of $m^{-1}$. This makes the dimensionality of $\boldsymbol{M}$ equal to that of $\mu_0$ (i.e., henry/meter) times the dimensions of $I\mathcal{A}\delta(x)\delta(y)\delta(z)$, which amount to ampere/meter. The latter, of course, are the dimensions of the $H$-field.